\pdfoutput=1

\documentclass[11pt]{article}

\usepackage{xcolor}
\usepackage{booktabs}

\usepackage{acl}
\usepackage{times}
\usepackage{latexsym}

\usepackage{graphicx}
\usepackage{caption}
\usepackage{changepage}
\usepackage{booktabs}

\usepackage[T1]{fontenc}

\usepackage[utf8]{inputenc}

\usepackage{microtype}

\usepackage{inconsolata}
\title{Scaling Automatic Extraction of Pseudocode}

\usepackage{makecell}

\author{%
  Levent Toksoz$^{1}$ \quad Gang Tan$^{1}$ \quad C. Lee Giles$^{1,2}$ \\
  \\
  \begin{tabular}{c@{\hskip 1.5cm}c}
    $^{1}$\makecell{The Pennsylvania State University, \\ Computer Science and Engineering, \\ University Park, Pennsylvania, 16802} & 
    $^{2}$\makecell{The Pennsylvania State University, \\ Information Sciences and Technology, \\ University Park, Pennsylvania, 16802} \\
  \end{tabular} \\
  \texttt{lkt5297@psu.edu} \quad \texttt{gtan@psu.edu} \quad \texttt{clg20@psu.edu}
}

\begin{document}
\maketitle

\begin{abstract}
Pseudocode in a scholarly paper provides a concise way to express the algorithms implemented therein. Pseudocode can also be thought of as an intermediary representation that helps bridge the gap between programming languages and natural languages. 
 Having access to a large collection of pseudocode can provide
various benefits ranging from enhancing algorithmic understanding, facilitating further algorithmic design, to empowering NLP or computer vision based models for tasks such as automated code generation and optical character recognition (OCR). We have created a large pseudocode collection by extracting nearly 320,000 pseudocode examples from arXiv papers. This process involved scanning over $2.2$ million scholarly papers, with 1,000 of them being manually inspected and labeled. Our approach  encompasses an extraction mechanism tailored to optimize the coverage and a validation mechanism based on random sampling to check its accuracy and reliability, given the inherent heterogeneity of the collection. In addition, we offer insights into common pseudocode structures, supported by clustering and statistical analyses. Notably, these analyses indicate an exponential-like growth in the usage of pseudocodes, highlighting their increasing significance.
\end{abstract}

\section{Introduction}

Pseudocode serves as an instrumental device, employed to express algorithms in a concise, syntactical constraint free format. Pseudocode also incorporates elements from both programming languages and natural languages, making it an ideal candidate to be used as an intermediate representation, bridging the gap between these languages. The elements incorporated from programming languages are often represented as universally recognized constructs employed throughout contemporary programming paradigms, thus highlighting the versatility of pseudocode and solidifying its role as a bridge between programming languages and natural languages. Given the importance of pseudocode, a large collection of pseudocode can enhance algorithmic understanding, facilitate algorithmic design, and empower NLP or computer vision based models for tasks such as automated code generation and optical character recognition (OCR). Some published works have explored the role of pseudocode in automated code generation tasks. A brief overview of these articles is presented in the related work section. 

A notable example of a pseudocode dataset is the SPOC dataset \citet{DBLP:journals/corr/abs-1906-04908} 
with approximately 20,000 pseudocodes along with their implementations and test cases.
The pseudocodes in the SPOC dataset were manually written by programmers contracted through Amazon Mechanical Turk. As such this dataset captures a limited scope of pseudocode examples, in stark contrast to the diverse range found in research papers such as those in arXiv,  which ranges from high-level pseudocodes to ones resembling actual code. It should be noted that the main purpose of the SPOC dataset is to provide a training/testing data-set for pseudocode to code conversion. Other notable examples include datasets from \citet{Oda2015LearningTG} and \citet{zavershynskyi2018naps}. The dataset provided by \citet{Oda2015LearningTG} offers around 16,000 manually written pseudocodes for statistical machine translation, and the one from \citet{zavershynskyi2018naps} provides approximately 2,000 manually written pseudocodes for the program synthesis task.

Our intent is to create a large collection of pseudocode extracted from arXiv papers, encompassing a diverse spectrum of pseudocode representations. Thus our collection can be employed for a variety of tasks. For instance, it can be used as a benchmark for automated code generation models and provide training/testing data for bimodal machine learning models to extract information like text and figures from PDFs by pairing the pseudocode with the texts describing the pseudocode in the PDFs. 

One challenge of extracting pseudocode from research papers is the heterogeneity of the files of research papers. Some research papers from arXiv have LaTex files in addition to PDF files, making extracting pseudocode a relatively straightforward process by searching for LaTex commands that mark pseudocode. Some papers, however, are only PDF files. Extracting information such as texts/figures from PDF files is a challenging task that often involves various subtasks including detecting the boundaries of figures/texts and converting these detected pieces into the chosen format. Due to myriad ways to create PDF files, the text and figures within the PDF files lack a universal pattern. Thus, it is a tedious task for computational algorithm based solutions to deal with, motivating the development of machine learning based solutions to detect and extract text/figures from PDFs. A concise overview of these approaches can be found in the related work section. Given the interest in developing machine learning based tools to extract information from PDFs, the pseudocode collection and its extraction pipeline can be a valuable tool to further facilitate the development of such areas of research.

In summary, our contributions are: 
\begin{itemize}
    \item a method for finding and extracting pseudocode from arXiv papers,
    \item a large collection of nearly 320,000 pseudocodes extracted from these papers, accompanied by 1000 PDFs labeled by humans to determine the presence of pseudocode,
    \item analysis of the increasing growth of pseudocode in arXiv papers,
    \item clustering pseudocode based on topics.
\end{itemize}

\section{Related Work}

Pseudocode stands as a foundational concept in computer science, frequently used to articulate algorithms in a concise manner and serving as a link between natural languages and coding. Its instructional potential is explored in various articles, such as those by \citet{10.1145/3304221.3325581} and \citet{odisho2016teaching}. Pseudocode is also employed across multitude of tasks, showcasing its versatility in diverse applications. Articles like \citet{unipseudo} utilizes the universality of the pseudocode to obtain a binary code similarity measure. Additionally, the work by \citet{mishra-etal-2023-prompting} explores the use of pseudocode as a prompt for Large Language Models.

One of the most thoroughly investigated applications of pseudocode in scholarly papers is its use in translating pseudocode into code and generating pseudocode for a variety of software tasks. To that extent, \citet{DBLP:journals/corr/abs-1906-04908} presents a machine learning model that can translate pseudocode to C++ code and a SPOC dataset that contains roughly around 20000 programs along with their human-authored pseudocodes and test cases. Similar approaches are found in the works of \citet{Oda2015LearningTG} and \citet{zavershynskyi2018naps}.\citet{Oda2015LearningTG} focuses on pseudocode-based statistical machine translation, presenting a collection of approximately 16,000 manually crafted pseudocodes. On the other hand, \citet{zavershynskyi2018naps} addresses the program synthesis task, offering a dataset that includes roughly 2,000 manually written pseudocodes for this task.

Given the challenges associated with finding extensive pseudocode databases, much of the scholarly work focused on Pseudocode to Code generation relies heavily on the SPOC dataset introduced by \citet{DBLP:journals/corr/abs-1906-04908}. \citet{DBLP:journals/corr/abs-2005-05927} uses a hierarchical beam search method that concentrates on specific semantic and syntactic constraints inherent in a program to further improve the model of \citet{DBLP:journals/corr/abs-1906-04908}. \citet{10.5555/3524938.3525939} utilizes the compiler output to repair outputs generated by programs synthesis tasks on SPOC dataset \citet{DBLP:journals/corr/abs-1906-04908}. \citet{pmlr-v119-shi20a} and \citet{pmlr-v139-xie21f} adopts Transformer based model that undertakes pseudocode to code generation on a line-by-line basis, with the assistance of the SPOC dataset \citet{DBLP:journals/corr/abs-1906-04908}.

Concerning the generation of pseudocode from code, \citet{Yang2021FinegrainedPG}
combines code feature extraction with a transformer trained on the the SPOC dataset \citet{DBLP:journals/corr/abs-1906-04908} to generate pseudocode from  C++ and Python programs.\citet{Aloklaarticle} employs retrieval-based transformer trained on the SPOC dataset \citet{DBLP:journals/corr/abs-1906-04908} to generate pseudocode from  C++ and Python programs. Moreover, \citet{sontakke2023knowledge}
specializes in transferring the knowledge from trained code to a pseudocode model to other models that has no paralell data. 

Additionally, the generation of pseudocodes is not limited to code to pseudocode conversion task. Pseudocode can also be generated by extracting information from various sources, such as PDF documents. While existing works explore various methods for extracting different types of information from diverse origins, there exists a notable gap in the literature regarding the specific challenge of generating pseudocode datasets from PDFs, particularly those sourced from platforms like arXiv. Advancements in the field are shown by the work of \citet{DBLP:journals/corr/abs-2004-14356}, \citet{nassar2022tableformer}, \citet{9085944}, \citet{DBLP:journals/corr/abs-2003-08005},
\citet{10.1145/1065385.1065418},
\citet{houyufang2019acl},
\citet{houyufang2021eacl},
\citet{blecher2023nougat}, and tools such as \citet{GROBID}, which employ machine learning models to precisely extract information from PDFs. These efforts extend to capturing mathematical equations, tables, figures, and metadata. However, these models are not designed to handle the structure of a pseudocode content within documents.

The generation of pseudocode from code and tasks related to code-to-pseudocode conversion is frequently influenced by the models employed in tasks such as code synthesis, infilling, and docstring generation.\cite{https://doi.org/10.48550/arxiv.2002.08155} proposes CodeBert, a bimodal pre-trained model compatible with both programming languages (PL) and natural languages (NL). \citet{https://doi.org/10.48550/arxiv.2009.08366} modifies the BERT model for code, emphasizing inherent structure through structure-aware pre-training tasks based on data flow graphs. \citet{xu2022systematic} deploys a GPT-2 architecture, training across 12 programming languages for code synthesis. \cite{fried2023incoder} proposes unified generative model that can do left to right program synthesis and code infilling. \cite{https://doi.org/10.48550/arxiv.2102.04664} introduces a benchmark dataset with around 15k examples and baseline model for various code based machine learning tasks, including code to code translation. \citet{Rozire2021LeveragingAU}, \citet{https://doi.org/10.48550/arxiv.2006.03511}, and \citet{https://doi.org/10.48550/arxiv.2102.07492} utilize unsupervised learning techniques for code-to-code translation on monolingual programming language data, with each training batch transitioning from one programming language to another.

\section{Pseudocode Data}
\begin{figure}[h]
  \centering
  \includegraphics[width=0.5\linewidth]{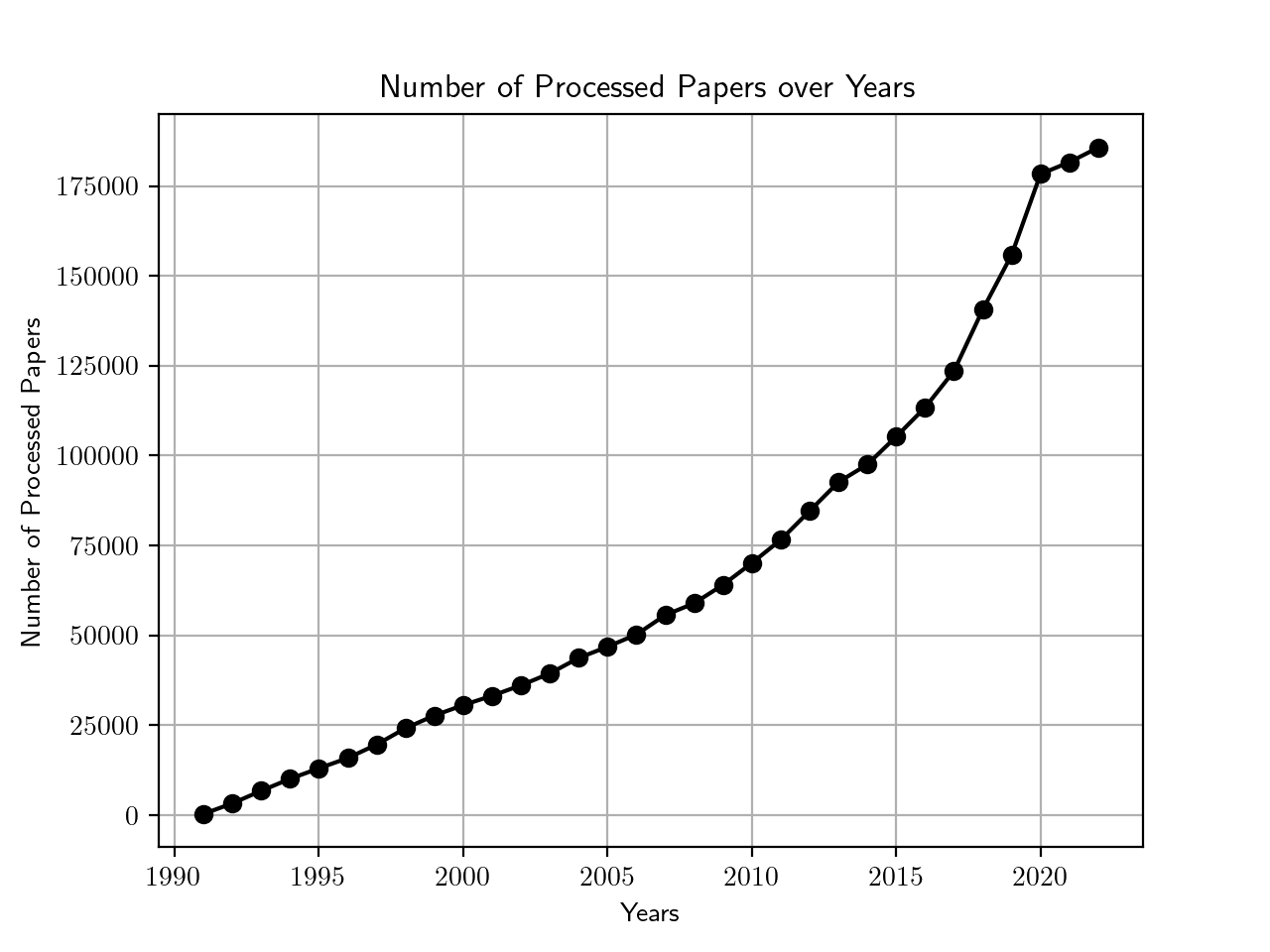}
  \caption{Number of papers scanned over years by the extraction pipeline.} 
  \label{fig:processed}
\end{figure}

\footnote{The dataset can be accessed via the \href{https://github.com/letoksoz/arxiv-pseudocode/}{\textit{arxiv-pseudocode}} repository on GitHub.}
Pseudocodes are extracted from the scholarly papers in arXiv. Approximately 10 TB of arXiv data is downloaded and stored across both Amazon S3 buckets and Google Cloud. In the arXiv dataset, there are 2.2 million PDF files, comprising roughly half of the total data size. The remaining data consists of supplementary files, such as images, figures, and LaTeX files. PDF files are retrieved using Google Cloud, while supplementary files like images, figures, and LaTeX files are extracted through Amazon S3 buckets. It should be noted that not all papers have associated supplementary files. The number of submissions to arXiv has been growing exponentially, underlining the significant contributions made in recent years as shown by the exponential-like trend depicted in Figure \ref{fig:processed}.

\section{Automatic Extraction}
To automatically detect and extract the pseudocode, we designed an extraction pipeline tailored to capture a diverse range of extracted pseudocode examples. Our pipeline has the following stages: preprocessing, pseudocode detection, and pseudocode extraction. It scans over $2.2$ million scholarly papers starting with year $1991$ and ending in June of $2023$. It should be noted that due to the large size of the dataset and the performance profile of the detection, extraction, and validation tools, running the pipeline requires a substantial amount of time, typically on the order of several days. 

\subsection{Preprocessing}
Some of the supplementary files in Amazon S3 are either stored as ZIP files, contain ZIP files, or even include both. Each ZIP file is extracted until no more ZIP files remain. Papers stored in arXiv has unique identifiers. By matching these unique identifiers, papers linked to Amazon S3 files (i.e., LaTeX files) are combined with their corresponding Google Cloud files (i.e., PDFs) as a single folder to be processed. 

\subsection{Pseudocode Detection}
To detect pseudocodes in each article, we first verify whether it contains LaTeX  files. If such a  file exists, we proceed to search for specific LaTex keywords: '\textbackslash begin\{algorithm\}' and '\textbackslash end\{algorithm\}' within the files. If an article's LaTeX files contain those keywords, it is forwarded to the extraction stage; otherwise, it is utilized to gather statistical information.


Some additional information about the papers are also stored, including 
PDF files themselves, supplementary files like LaTeX and HTML files, extracted text snippets from PDFbox \citet{pdfbox2012} and arXiv metadata. The arXiv metadata information includes the arXiv identifier of the paper, version of the PDF, the arXiv link, the abstract, the year in which it was uploaded to arXiv, the topic and subtopic of the article, and finally the title of the PDF. Our detection algorithm identified pseudocode in $141,939$ out of the $2.2$ million papers as shown in table \ref{tab:paperstats}.  

\subsection{Pseudocode Extraction}
In this stage, papers containing LaTeX files identified in the detection stage are processed. The pseudocode in these LaTeX files exhibit a heterogeneous structure due to the absence of a standardized LaTeX notation. For instance, \textbackslash {begin\{equation\}} and \textbackslash {begin\{align\}} might serve the same purpose. Therefore, it might be unrealistic to expect the extraction algorithm to cover all possible representations. As we will explain further in the validation and statistical analyses section, utilizing \textbackslash{begin\{algorithm\}} tags to detect and extract pseudocode from LaTeX papers is reliable. Our analyses also indicate that the majority of the papers in arXiv have LaTeX files, underlining the significance of extracting information from LaTex files. For that reason, the extraction algorithm relies on \textbackslash{begin\{algorithm\}} and \textbackslash{end\{algorithm\}} tags. Specifically, our algorithm identifies the locations of those tags and then extracts the pseudocode found between them. When pseudocode includes references to other LaTex contents such as equations, we search for the corresponding label of that reference within the file and then extract it as supplementary information. 
Our extraction mechanism obtained $323,303$ pseudocodes and saved each as a JSON file, along with metadata information such as the arXiv identifier, any equations referenced by the pseudocode, and the year it was stored in arXiv. 

\section{Validation}
The validation aims to understand the accuracy of our pseudocode exaction mechanism. A false positive is a paper that does not contain pseudocode but our extraction extracted pseudocode from it.
A false negative is a paper that does contain pseudocode but our exctraction cannot extract pseudocode.

Due to the extensive size and characteristics of the dataset, it is not feasible to obtain the ground truth such as whether a paper contains pseudocode or not for all the papers in the dataset.
To that extent, we utilized a sampling based approach. We uniformly sampled $1000$ PDFs among all the scanned papers spanning from the year $1991$ to $2023$. Each sampled paper was then manually inspected to determine whether it contains pseudocode and labeled accordingly. The distribution of pseudocode counts within these inspected papers is shown in Table \ref{tab:inspectedpaper}. In addition to labeling each pseudocode, we store additional metadata about the paper.

\begin{table}
    \centering
    \caption{Sampled Counts}
    \label{tab:inspectedpaper}
    \begin{tabular}{lc}
    \hline
    \textbf{Manually Inspected 1000 Papers} & \textbf{Number} \\
    \hline
    Has Pseudocode & $101$ \\
    Does not have Pseudocode & $899$ \\ 
    \hline
    \end{tabular}
\end{table}

\begin{table}
    \centering
    \caption{False Positive and False Negative Rates}
    \label{tab:falserates}
    \begin{tabular}{lc}
    \hline
    \textbf{Type} & \textbf{ Percentage } \\
    \hline
    FPR & $\% 0.6$ \\
    FNR & $\%33.7$ \\ 
    \hline
    \end{tabular}
\end{table}

By cross-checking the results obtained from our detection-extraction mechanism with this manually labeled sampled set, we computed the false negative and false positive rates presented in Table \ref{tab:falserates}. These results indicate that the extraction-detection mechanism, based on the '\textbackslash{begin\{algorithm}\}' tag, is a reliable method for detecting and extracting pseudocodes in LaTeX. However, it could be improved, as it sometimes misses certain pseudocode patterns. Upon manual inspection of some of these overlooked patterns, we observed that there were manually structured pseudocodes with '\textbackslash{begin\{enumerate}\}' tags and other pseudocodes embedded within paragraphs. Some papers also utilized '\textbackslash{begin\{algorithm\}}' to set up problems rather than pseudocode descriptions. However, the overall number was very low within the sampled set, contributing to the low false positive rate of our extraction process.

\section{Statistical Analyses}

\begin{table}
    \centering
    \caption{Paper Counts}
    \label{tab:papercounts}
    \begin{tabular}{lc}
    \hline
    \textbf{From year 1991 to year 2023} & \textbf{Number} \\
    \hline
    Total papers & $2,285,111$ \\
    Papers with LaTex & $2,054,422$ \\ 
    \hline
    \end{tabular}
\end{table}

\begin{table}
    \centering
    \caption{Papers with pseudocode}
    \label{tab:paperstats}
    \begin{tabular}{lc}
    \hline
    \textbf{From year 1991 to year 2023} & \textbf{Number} \\ 
    \hline
    Papers with keywords indicating the \\ presence of pseudocode & $241,275$ \\
    Papers with tag \\ "\textbackslash{begin\{algorithm\}"} & $141,939$ \\ \hline
    \end{tabular}
\end{table}

Results of our extraction mechanism are briefly summarized by Table~\ref{tab:papercounts}, Figure \ref{fig:processed}, and Figure \ref{fig:pseudo}. Significantly, around $90\%$ of the papers were accompanied by LaTeX files, reinforcing the choice to extract pseudocodes from LaTeX files instead of PDFs. Another notable result is the number of papers with pseudocode grows almost exponentially over the years, as shown by Figure \ref{fig:pseudo}, underlining the increasing significance of pseudocode. We hypothesize that this increasing trend correlates with the growing prominence of computer science subjects in arXiv, as well as the increased availability of powerful computing tools and growing computational capacity.

\begin{figure}[h]
    \centering   \includegraphics[width=0.5\linewidth]{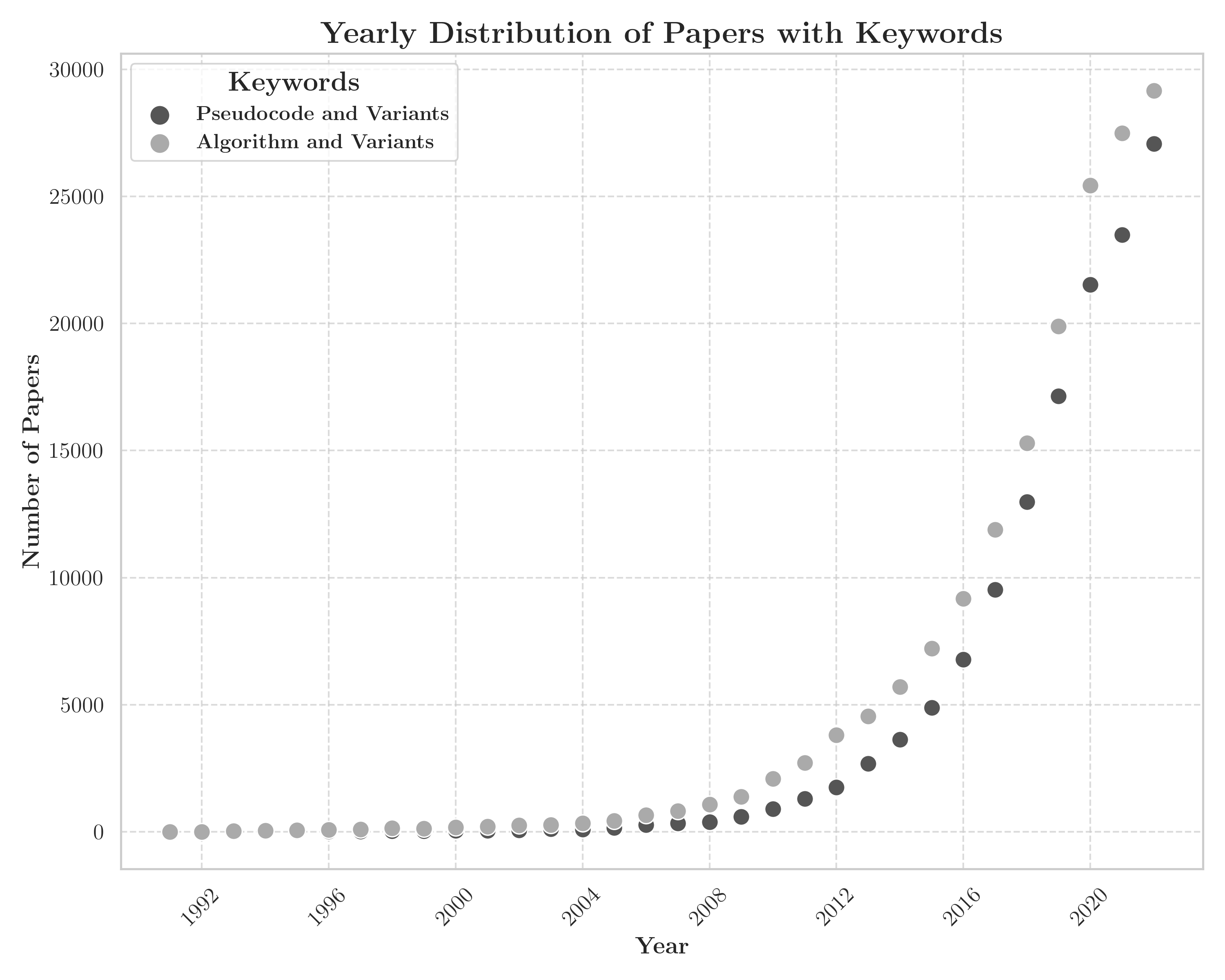}
    \caption{Yearly Distribution of Papers with Keywords (See Table~\ref{tab:indicword} for keyword details)}
    \label{fig:yearly_distribution}
\end{figure}
\begin{figure*}[h]
    \centering   \includegraphics[width=0.8\linewidth]{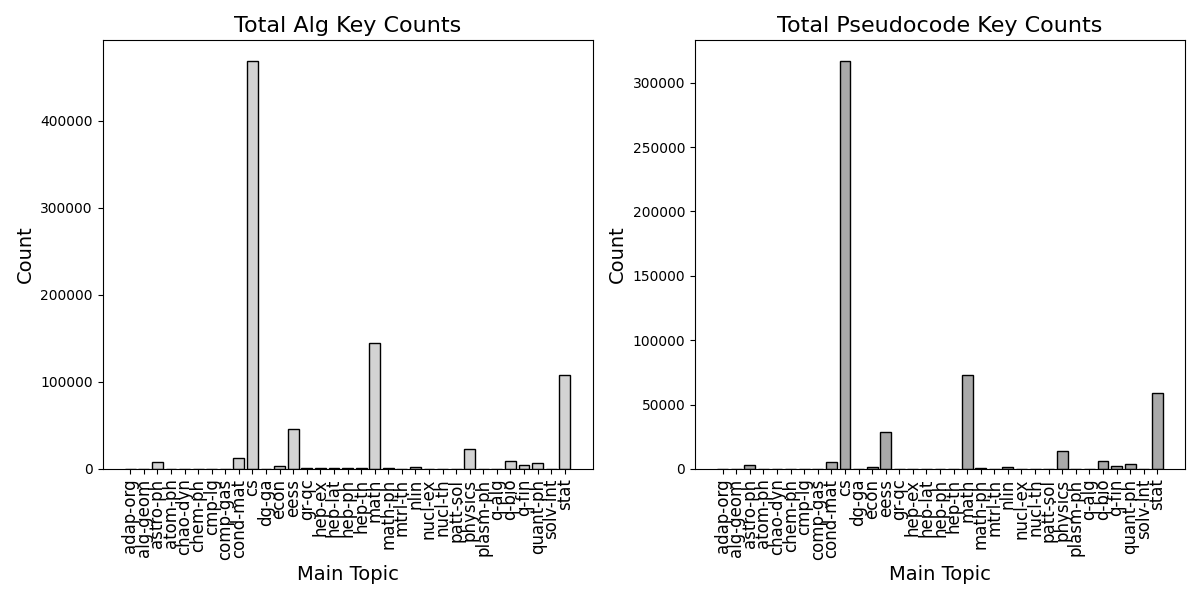}
    \caption{Category Distribution of Papers with Keywords (See Table~\ref{tab:indicword} for keyword details)}
    \label{fig:cat_distribution}
\end{figure*}

We also analyzed the papers that use keywords indicative of the presence of pseudocode, such as "Algorithm $n$", where $n$ is a number, and "Pseudocode". Full list of words can be found in Table \ref{tab:indicword}. Note that given the popularity of the word "Algorithm", the algorithm keyword alone does not directly correlate with the presence of a pseudocode.
To identify such papers, we perform a direct search within the PDF files using Apache PDFBox \citet{pdfbox2012}. For the corresponding LaTeX source code, similar methodology mentioned in Pseudocode detection section is employed. PDFBox \citet{pdfbox2012} extracts various information from the PDF, including text, formatting options, and metadata about the article. We conduct a search within the extracted text of PDF files for specific keywords such as 'Pseudocode' and 'Algorithm 1'. The number distribution of these papers are shown by Table \ref{tab:paperstats} and figure \ref{fig:yearly_distribution}.Both keyword paper counts exhibit exponential-like growth. To validate whether the presence of these indicative words indicates the presence of pseudocode, we used our 1000
manually examined validation paper set. The results are shown in Table \ref{tab:indicativepaper}. It shows that the indicative words alone are relatively insufficient to reliably indicate the presence of pseudocode given their relatively large false positive count compared to our mechanism. However, false negative count decreases, enhancing the detection of various pseudocode types.  
Using the corresponding metadata information of each PDF that contains these indicative words, a category plot for each document is displayed in Figure \ref{fig:cat_distribution}, indicating that most of these keywords occur in computer science-themed papers.

\begin{table}
    \centering
    \caption{Indicative Words.}
    \label{tab:indicword}
    \begin{tabular}{|l|p{12cm}|} 
    \hline
    \textbf{Keyword} & \textbf{Searched Words} \\
    \hline
    Pseudocode & "Pseudocode", "pseudocode", "Pseudo-code", "pseudo-code" \\
    Algorithm & "Algorithm N", "algorithm N", "Algorithm-N", "algorithm-N", "Algorithm:", "algorithm:" \\
    \hline
    \end{tabular}
\end{table}

\begin{table}
    \centering
    \caption{Indicative Word Inspection.}
    \label{tab:indicativepaper}
    \begin{tabular}{lccc}
    \hline
    \textbf{Has Pseudocode} & \textbf{Sampled 1000 Papers} & \textbf{Contains Indicative Keywords} \\
    \hline
    Yes & $101$ & $75$ \\
    No & $899$ & $20$ \\ 
    \hline
    \end{tabular}
\end{table}

\begin{figure}[h]
  \centering
  \label{fig:pseudo}
  \includegraphics[width=0.5\linewidth]{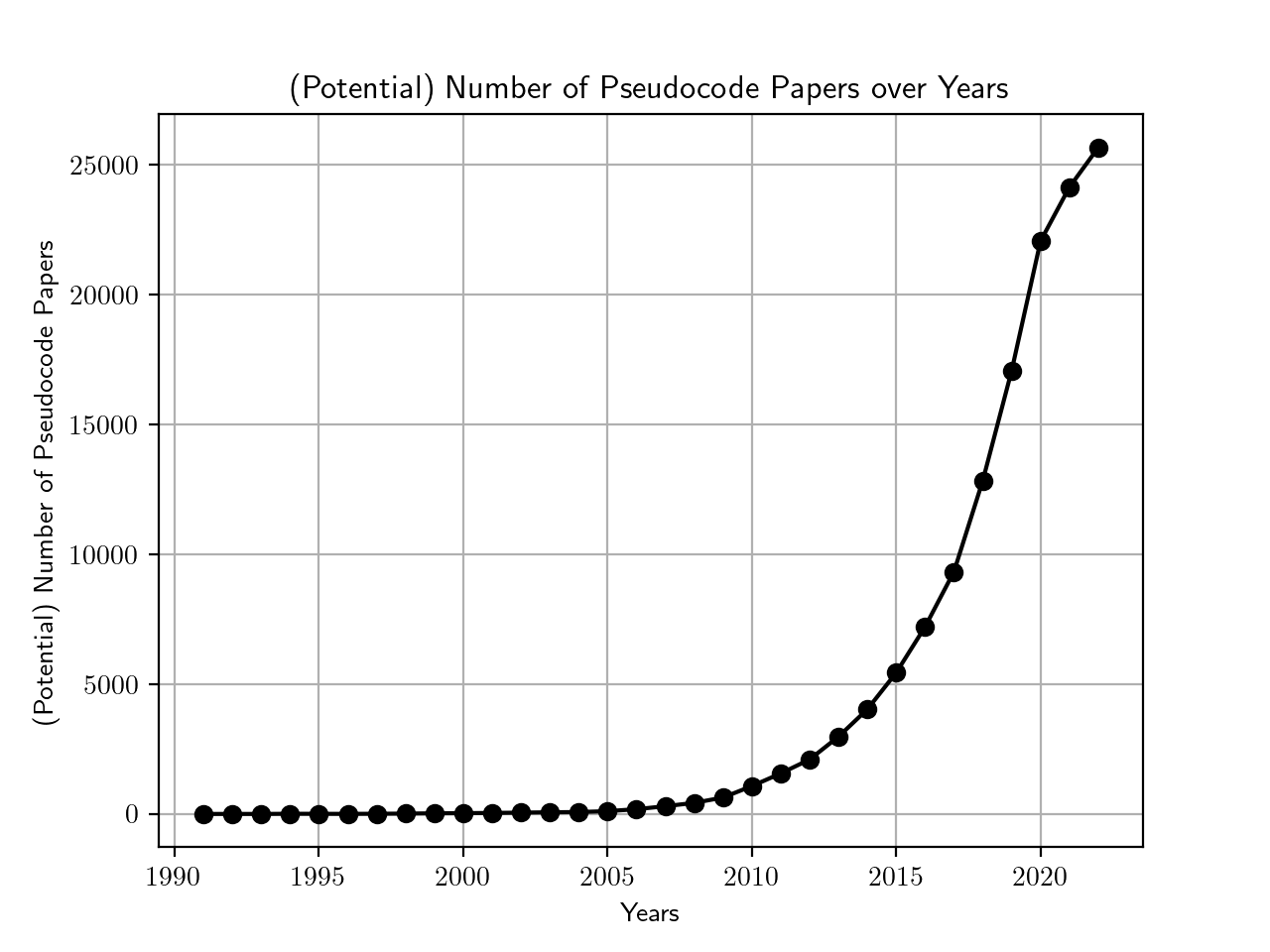}
  \caption{Number of papers with LaTeX and "\textbackslash {begin\{algorithm\}"} tag.}
  \label{fig:pseudo}
\end{figure}

\section{Clustering}

We investigate how the topics of pseudocodes evolve over time by clustering. Using the arXiv metadata, the topics of each paper containing pseudocode can be plotted, similar to Figure \ref{fig:cat_distribution}. However, arXiv topics are often too broad and do not precisely represent the pseudocode topics. For instance, a biology-themed paper may include pseudocode related to computer science or graph theory. To address this, we have designed our own clustering mechanism based on the text snippets where pseudocode is mentioned.

The clustering mechanism utilizes text snippets that reference the pseudocode. These snippets are created as a result of our reference detection mechanism and are cleaned from irrelevant LaTeX syntax. Since the number of papers containing pseudocode before the year $2010$ is negligible, as indicated in Figure \ref{fig:pseudo}, we exclusively utilized text references from the year $2010$ onward. Additionally, common English stop words and non-instructive words for topic modeling, such as 'use', 'employ', and 'indicate' are filtered out. To consolidate different variations of the same word, such as 'decode,' 'decoding,' and 'decoded,' into a single representation, each word undergoes stemming. The Term Frequency-Inverse Document Frequency Vectorizer (TF-IDF) is utilized to obtain vector representations for each text snippet. Terms that appear in more than $85$ percent and less than $0.02$ percent of documents are disregarded.

\subsection{Reference Detection Algorithm}

The reference detection mechanism operates as follows: Whenever pseudocode includes a label, our mechanism attempts to locate that label. Given that labels can be represented in various ways in LaTeX, we employ regular expressions (regex) to accommodate the diverse representations of labels, which also account for special characters. Once a label is found, the mechanism dynamically generates the associated reference tag. These reference tags are tailored to different types of reference words (e.g., equations, algorithms, theorems, etc.), which typically have an identifying segment followed by 'ref' (e.g., algref, eqref). To handle the variations specific to each type of reference, we employ regular expressions (regex) for pattern matching. The generated references are then searched within the entire directory, as these references could potentially be located in other LaTeX files. Finally, when a reference tag is identified, we mark a span of $1200$ characters before and after that tag. If the $300$-character span around these marked locations contains a character indicating the end of a sentence, we extract the text up to that point. However, if no such characters are found, we retain the original span of $1200$ characters. Our mechanism successfully extracted the references of the majority of pseudocodes. The ones that were not extracted either lack associated references or possess a reference tag that is too complex for regular expressions to handle.

\subsection{Cleaning Extracted Text}

The extracted reference texts for each pseudocode are presented in LaTeX format, containing LaTeX keywords, special symbols, comments, and mathematical symbols. To clean the extracted reference texts, removal processes are implemented for different LaTeX elements: LaTeX keywords by detecting \textbackslash, comments via \%, mathematical symbols through \$, and special words containing a single character on either side of \_ (or both sides or none). 

\subsection{Topic Modelling}

\begin{table}
    \caption{Pseudocode Topic Clusters}
    \label{tab:topiccluster}
    \tiny 
    \begin{minipage}[t]{0.45\linewidth}
    \begin{tabular}{ccl}
    \toprule
    \textbf{Years} & \textbf{Cluster Number} & \textbf{Top $5$ words represented in stemmed format}\\ 
    \midrule

    $2010$ & $0$ & polici, stage, edg, algorithms, channel\\
    $2010$ & $1$ & messag, node, root, max, tree\\
    $2010$ & $2$ & \textcolor{red}{particl, posterior, filter, sampl, likelihood}\\
    $2010$ & $3$ & protocol, request, receiv, node, neighbor\\
    $2010$ & $4$ & bit, specif, read, string, regular\\
    $2010$ & $5$ & matrix, optim, iter, vector, estim\\
    $2010$ & $6$ & \textbf{node, edg, graph, polynomi, vertex}\\
    $2010$ & $7$ & threshold, delet, reconstruct, spars, recurs\\
    $2010$ & $8$ & insid, bit, residu, embed, exist\\
    $2010$ & $9$ & thread, processor, reconstruct, queue, termin\\ 

    \hline
    $2011$ & $0$ & cluster, read, object, link, return\\
    $2011$ & $1$ & \textcolor{red}{atom, support, hard, potenti, previou}\\
    $2011$ & $2$ & decod, eq, nearest, eigenvector, neighbor\\
    $2011$ & $3$ & node, protocol, neighbor, messag, cell\\
    $2011$ & $4$ & pseudocod, present, scheme, accord, symbol\\
    $2011$ & $5$ & game, strategi, word, compress, alg\\
    $2011$ & $6$ & tabl, tupl, densiti, sampl, length\\
    $2011$ & $7$ & \textbf{graph, node, edg, bound, path}\\
    $2011$ & $8$ & matrix, iter, converg, vector, column\\
    $2011$ & $9$ & respons, estim, disjoint, prefix, wait\\ 

    \hline
    $2012$ & $0$ & tree, cell, root, univers, parent\\
    $2012$ & $1$ & score, box, algorithms, symbol, train\\
    $2012$ & $2$ & matrix, iter, optim, converg, approxim\\
    $2012$ & $3$ & state, block, messag, node, return\\
    $2012$ & $4$ & \textbf{edg, graph, vertex, vertic, node}\\
    $2012$ & $5$ & aggreg, domin, parallel, classifi, join\\
    $2012$ & $6$ & page, environ, polici, qualiti, regress\\
    $2012$ & $7$ & prime, ideal, tupl, ration, relationship\\
    $2012$ & $8$ & \textcolor{red}{node, sampl, cluster, particl, optim}\\
    $2012$ & $9$ & constraint, equation, pseudocod, add, achiev\\ 
    
    \hline
    $2013$ & $0$ & node, probabl, round, random, strategi\\
    $2013$ & $1$ & \textcolor{red}{sampl, distribut, posterior, particl, densiti}\\
    $2013$ & $2$ & messag, protocol, key, decod, receiv\\
    $2013$ & $3$ & box, array, cell, block, memori\\
    $2013$ & $4$ & column, row, read, decod, symbol\\
    $2013$ & $5$ & user, role, request, server, polici\\
    $2013$ & $6$ & cluster, segment, flow, label, chang\\
    $2013$ & $7$ & matrix, iter, converg, optim, estim\\
    $2013$ & $8$ & \textbf{edg, node, graph, tree, path}\\
    $2013$ & $9$ & node, agent, network, commun, alloc\\ 

    \hline
    $2014$ & $0$ & decod, code, encod, array, bit\\
    $2014$ & $1$ & matrix, iter, converg, vector, optim\\
    $2014$ & $2$ & alloc, resourc, power, rate, user\\
    $2014$ & $3$ & \textcolor{red}{sampl, distribut, estim, particl, posterior}\\
    $2014$ & $4$ & cell, refin, mesh, trace, insert \\
    $2014$ & $5$ & \textbf{node, edg, graph, path, tree}\\
    $2014$ & $6$ & \textcolor{blue}{featur, train, classifi, layer, classif}\\
    $2014$ & $7$ & mechan, mesh, grid, truth, increment\\
    $2014$ & $8$ & item, user, price, messag, alloc \\
    $2014$ & $9$ & label, pattern, state, block, end\\ 

    \hline
    
    $2015$ & $0$ & \textcolor{blue}{cluster, train, featur, label, learn}\\
    $2015$ & $1$ & cell, row, array, block, matrix\\
    $2015$ & $2$ & \textbf{node, graph, edg, tree, path}\\
    $2015$ & $3$ & optim, constraint, alloc, feasibl, power\\
    $2015$ & $4$ & \textcolor{blue}{layer, gradient, batch, train, learn} \\
    $2015$ & $5$ & central, author, walk, algorithms, contribut\\
    $2015$ & $6$ & polynomi, theorem, element, bound, return\\
    $2015$ & $7$ & agent, node, messag, protocol, round\\
    $2015$ & $8$ & converg, iter, convex, optim, solv \\
    $2015$ & $9$ & sampl, matrix, estim, distribut, approxim\\

    \hline
    $2016$ & $0$ & polynomi, satisfi, state, return, theorem\\
    $2016$ & $1$ & optim, iter, converg, solv, gradient\\
    $2016$ & $2$ & \textbf{node, edg, graph, vertex, vertic}\\
    $2016$ & $3$ & block, array, item, parallel, cell\\
    $2016$ & $4$ & cluster, frame, assign, merg, distanc \\
    $2016$ & $5$ & agent, action, state, transit, flow\\
    $2016$ & $6$ & matrix, probabl, rank, sampl, column\\
    $2016$ & $7$ & estim, sampl, matrix, distribut, approxim\\
    $2016$ & $8$ & layer, pixel, bit, encod, code\\
    $2016$ & $9$ & \textcolor{blue}{train, learn, featur, user, network}\\

        \\\bottomrule

    \end{tabular}
\end{minipage}
  \hspace{.05\linewidth}
  \begin{minipage}[t]{.45\linewidth}
    \begin{tabular}{ccl}
    \toprule
    \textbf{Years} & \textbf{Cluster Number} & \textbf{Top $5$ words represented in stemmed format}\\ 
    \midrule

    \hline
    $2017$ & $0$ & estim, sampl, distribut, approxim, simul\\
    $2017$ & $1$ & \textbf{node, tree, edg, graph, path}\\
    $2017$ & $2$ & cluster, color, schedul, node, flow\\
    $2017$ & $3$ & agent, polici, action, state, reward\\
    $2017$ & $4$ & iter, converg, bound, theorem, optim \\
    $2017$ & $5$ & matrix, sampl, tensor, distribut, row\\
    $2017$ & $6$ & user, bit, node, code, protocol\\
    $2017$ & $7$ & \textbf{edg, graph, vertic, vertex, element}\\
    $2017$ & $8$ & cell, item, alloc, budget, greedi\\
    $2017$ & $9$ & \textcolor{blue}{train, learn, layer, gradient, network}\\

    \hline
    $2018$ & $0$ & \textcolor{blue}{featur, label, train, class, classifi}\\
    $2018$ & $1$ & \textbf{node, tree, path, graph, edg}\\
    $2018$ & $2$ & solv, iter, optim, constraint, converg\\
    $2018$ & $3$ & user, optim, alloc, power, channel\\
    $2018$ & $4$ & cluster, item, popul, cell, individu \\
    $2018$ & $5$ & matrix, estim, sampl, approxim, distribut\\
    $2018$ & $6$ & bound, theorem, sampl, converg, gradient\\
    $2018$ & $7$ & \textcolor{blue}{train, learn, network, layer, gradient}\\
    $2018$ & $8$ & \textbf{edg, graph, vertex, vertic, block}\\
    $2018$ & $9$ & agent, action, state, polici, reward\\
    \hline
    $2019$ & $0$ & matrix, vector, matric, rank, approxim\\
    $2019$ & $1$ & polici, action, agent, state, reward\\
    $2019$ & $2$ & attack, adversari, test, perturb, detect\\
    $2019$ & $3$ & regret, bound, gradient, theorem, sgd\\
    $2019$ & $4$ & \textcolor{blue}{train, network, layer, learn, featur} \\
    $2019$ & $5$ & node, item, queri, messag, block\\
    $2019$ & $6$ & iter, converg, optim, solv, convex\\
    $2019$ & $7$ & \textbf{node, edg, graph, path, vertex}\\
    $2019$ & $8$ & sampl, estim, distribut, posterior, probabl\\
    $2019$ & $9$ & cluster, tree, cell, node, partit\\
    
    \hline
    
    $2020$ & $0$ & iter, converg, matrix, bound, theorem\\
    $2020$ & $1$ & polici, action, agent, state, reward\\
    $2020$ & $2$ & bound, queri, probabl, theorem, return\\
    $2020$ & $3$ & optim, user, alloc, power, channel\\
    $2020$ & $4$ & messag, block, protocol, server, bit\\
    $2020$ & $5$ & \textcolor{blue}{train, featur, layer, label, predict}\\
    $2020$ & $6$ & \textbf{node, edg, graph, tree, path}\\
    $2020$ & $7$ & solv, iter, optim, equat, constraint\\
    $2020$ & $8$ & sampl, estim, distribut, arm, posterior\\
    $2020$ & $9$ & \textcolor{blue}{train, loss, attack, learn, gradient}\\ 

    \hline
    $2021$ & $0$ & sampl, estim, distribut, posterior, gaussian\\
    $2021$ & $1$ & \textbf{node, edg, graph, tree, path}\\
    $2021$ & $2$ & cluster, test, estim, popul, error\\
    $2021$ & $3$ & \textcolor{blue}{train, loss, learn, network, layer}\\
    $2021$ & $4$ & environ, packag, column, tabl, includ\\
    $2021$ & $5$ & client, server, messag, protocol, local\\
    $2021$ & $6$ & matrix, iter, solv, vector, optim\\
    $2021$ & $7$ & bound, theorem, converg, iter, proof\\
    $2021$ & $8$ & polici, action, agent, state, reward\\
    $2021$ & $9$ & alloc, user, item, schedul, optim\\ 

    \hline
    $2022$ & $0$ & polici, agent, action, state, reward\\
    $2022$ & $1$ & optim, iter, solv, converg, constraint\\
    $2022$ & $2$ & probabl, theorem, bound, return, satisfi\\
    $2022$ & $3$ & sampl, distribut, estim, posterior, probabl\\
    $2022$ & $4$ & attack, score, user, detect, adversari\\
    $2022$ & $5$ & iter, cell, optim, perform, fig\\
    $2022$ & $6$ & matrix, vector, matric, rank, tensor\\
    $2022$ & $7$ & \textcolor{blue}{train, loss, learn, layer, featur}\\
    $2022$ & $8$ & bound, regret, theorem, arm, estim\\
    $2022$ & $9$ & \textbf{node, edg, graph, tree, path}\\ 

    \hline
    $2023$ & $0$ & attack, adversari, game, perturb, box\\
    $2023$ & $1$ & client, train, server, featur, dataset\\
    $2023$ & $2$ & iter, converg, solv, optim, matrix\\
    $2023$ & $3$ & \textbf{edg, node, graph, vertex, vertic}\\
    $2023$ & $4$ & \textcolor{blue}{train, loss, sampl, learn, network}\\
    $2023$ & $5$ & sampl, estim, distribut, simul, probabl\\
    $2023$ & $6$ & path, robot, search, state, optim\\
    $2023$ & $7$ & regret, bound, proof, theorem, reward\\
    $2023$ & $8$ & polici, agent, action, reward, state\\
    $2023$ & $9$ & node, cluster, matrix, oper, block\\ 
    \\\bottomrule

    \end{tabular}
    \end{minipage}
\end{table}

To investigate how the topics of pseudocodes evolve over time, we employed Latent Dirichlet Allocation (LDA) to perform soft clustering of the data, allowing each pseudocode reference to have multiple topic representations. To obtain consistency across years, the number of topics is set to $10$, which might not be optimal for certain years given the significant variation in the number of pseudocodes across years as shown in figure \ref{tab:topiccluster}. To understand the representation of each topic, we provide the top 5 most representative words for each topic, listed in descending order of relevance. The clustering results are displayed in Table \ref{tab:topiccluster}. From Table \ref{tab:topiccluster}, highlighted in bold, starting in $2010$ and continuing thereafter, there is a noticeable emphasis on pseudocodes related to graph algorithms. We can also observe a diminishing prominence of physics-related pseudocode. This transition can be observed from Table \ref{tab:topiccluster}, marked in red. Initially, physics-related words begin to be superseded by probability-related terms within the same cluster as years progress from $2010$ to $2014$. Furthermore, after the year $2014$, physics-related terms cease to appear. Starting from $2014$ and continuing up to $2023$, we observe machine learning to emerge as one of the prevailing topics for pseudocode, highlighted in blue in Table \ref{tab:topiccluster}.

\section{Conclusion}

We developed a pseudocode collection pipeline that we utilized to create a large collection of pseudocode examples, totaling approximately $320,000$. Due to the significant heterogeneity of the collection, we have built a sampling-based validation mechanism to ensure the reliability of our pipeline. Additionally, we manually inspected $1000$ papers and labeled them based on whether they contain pseudocode, along with their supplementary information. Moreover, we have employed clustering techniques, specifically LDA, to reveal thematic structures in a large collection of pseudocodes. Significantly, our findings reveal an exponential growth in the utilization of pseudocodes over time, with a particular focus on pseudocodes related to graph algorithms emerging as a predominant theme.

\section{Future Work}

Our dataset has the potential to serve as a valuable resource for a wide range of applications such as empowering NLP or computer vision based models and enhancing algorithmic understanding, opening up avenues for diverse future research. In this section, we will briefly discuss how our dataset can be utilized for these potential applications. 

In our dataset, each pseudocode is linked to its respective arXiv identifier. Leveraging these identifiers, we can establish connections between LaTeX-formatted pseudocodes and their corresponding PDFs. By using these pairs as training and testing data for bimodal machine learning models, we can partially automate the extraction of information, including text and figures, from the PDFs. Additionally, despite the prevalence of LaTeX files, we have also identified around $25,000$ papers without LaTeX files. These papers are potential candidates that may contain pseudocode and could greatly benefit from such an automated process. Moreover, we manually inspected $1000$ papers and labeled them
based on whether they contain pseudocode, along with their supplementary information.

There are other valuable uses of our dataset. If we convert our LaTeX-formatted pseudocodes into a more text-like format, they could serve as testing and benchmarking data for automated code generation tasks. This can be achieved to some extent using tools like \citet{latexml}. Such tasks involve converting pseudocode to actual code or utilizing pseudocode as an intermediary representation for natural language to code or code to natural language translation.

Another interesting project would to be to build a focused search that permits the indexing and searching of pseudocode in order to facilitate the use and discovery of related pseudocode.

\section{Acknowledgements}
The arXiv is gratefully acknowledged for providing access to documents with pseudocode and their latex versions.

\vspace{15mm}



\begin{thebibliography}{33}
\expandafter\ifx\csname natexlab\endcsname\relax\def\natexlab#1{#1}\fi

\bibitem[{GRO(2008--2023)}]{GROBID}
 2008--2023.
\newblock \href {http://arxiv.org/abs/1:dir:dab86b296e3c3216e2241968f0d63b68e8209d3c} {Grobid}.
\newblock \url{https://github.com/kermitt2/grobid}.

\bibitem[{Alokla et~al.(2022)Alokla, Gad, Nazih, Aref, and M.Salem}]{Aloklaarticle}
Anas Alokla, Walaa Gad, Waleed Nazih, Mostafa Aref, and Abdel-Badeeh M.Salem. 2022.
\newblock \href {https://doi.org/10.3390/math10040604} {Retrieval-based transformer pseudocode generation}.
\newblock \emph{Mathematics}, 10:604.

\bibitem[{Blecher et~al.(2023)Blecher, Cucurull, Scialom, and Stojnic}]{blecher2023nougat}
Lukas Blecher, Guillem Cucurull, Thomas Scialom, and Robert Stojnic. 2023.
\newblock \href {http://arxiv.org/abs/2308.13418} {Nougat: Neural optical understanding for academic documents}.

\bibitem[{Davila et~al.(2021)Davila, Setlur, Doermann, Kota, and Govindaraju}]{9085944}
Kenny Davila, Srirangaraj Setlur, David Doermann, Bhargava~Urala Kota, and Venu Govindaraju. 2021.
\newblock \href {https://doi.org/10.1109/TPAMI.2020.2992028} {Chart mining: A survey of methods for automated chart analysis}.
\newblock \emph{IEEE Transactions on Pattern Analysis and Machine Intelligence}, 43(11):3799--3819.

\bibitem[{Feng et~al.(2020)Feng, Guo, Tang, Duan, Feng, Gong, Shou, Qin, Liu, Jiang, and Zhou}]{https://doi.org/10.48550/arxiv.2002.08155}
Zhangyin Feng, Daya Guo, Duyu Tang, Nan Duan, Xiaocheng Feng, Ming Gong, Linjun Shou, Bing Qin, Ting Liu, Daxin Jiang, and Ming Zhou. 2020.
\newblock \href {https://doi.org/10.48550/ARXIV.2002.08155} {Codebert: A pre-trained model for programming and natural languages}.

\bibitem[{Fried et~al.(2023)Fried, Aghajanyan, Lin, Wang, Wallace, Shi, Zhong, tau Yih, Zettlemoyer, and Lewis}]{fried2023incoder}
Daniel Fried, Armen Aghajanyan, Jessy Lin, Sida Wang, Eric Wallace, Freda Shi, Ruiqi Zhong, Wen tau Yih, Luke Zettlemoyer, and Mike Lewis. 2023.
\newblock \href {http://arxiv.org/abs/2204.05999} {Incoder: A generative model for code infilling and synthesis}.

\bibitem[{Guo et~al.(2020)Guo, Ren, Lu, Feng, Tang, Liu, Zhou, Duan, Svyatkovskiy, Fu, Tufano, Deng, Clement, Drain, Sundaresan, Yin, Jiang, and Zhou}]{https://doi.org/10.48550/arxiv.2009.08366}
Daya Guo, Shuo Ren, Shuai Lu, Zhangyin Feng, Duyu Tang, Shujie Liu, Long Zhou, Nan Duan, Alexey Svyatkovskiy, Shengyu Fu, Michele Tufano, Shao~Kun Deng, Colin Clement, Dawn Drain, Neel Sundaresan, Jian Yin, Daxin Jiang, and Ming Zhou. 2020.
\newblock \href {https://doi.org/10.48550/ARXIV.2009.08366} {Graphcodebert: Pre-training code representations with data flow}.

\bibitem[{Hou et~al.(2019)Hou, Jochim, Gleize, Bonin, and Ganguly}]{houyufang2019acl}
Yufang Hou, Charles Jochim, Martin Gleize, Francesca Bonin, and Debasis Ganguly. 2019.
\newblock Identification of tasks, datasets, evaluation metrics, and numeric scores for scientific leaderboards construction.
\newblock In \emph{Proceedings of the 57th Annual Meeting of the Association for Computational Linguistics, {\em Florence, Italy, 27 July -- 2 August 2019}}.

\bibitem[{Hou et~al.(2021)Hou, Jochim, Gleize, Bonin, and Ganguly}]{houyufang2021eacl}
Yufang Hou, Charles Jochim, Martin Gleize, Francesca Bonin, and Debasis Ganguly. 2021.
\newblock Tdmsci: A specialized corpus for scientific literature entity tagging of tasks datasets and metrics.
\newblock In \emph{Proceedings of the the 16th conference of the European Chapter of the Association for Computational Linguistics, {\em Online, 19--23 April 2021}}.

\bibitem[{Hu et~al.(2005)Hu, Li, Cao, Meyerzon, and Zheng}]{10.1145/1065385.1065418}
Yunhua Hu, Hang Li, Yunbo Cao, Dmitriy Meyerzon, and Qinghua Zheng. 2005.
\newblock \href {https://doi.org/10.1145/1065385.1065418} {Automatic extraction of titles from general documents using machine learning}.
\newblock In \emph{Proceedings of the 5th ACM/IEEE-CS Joint Conference on Digital Libraries}, JCDL '05, page 145–154, New York, NY, USA. Association for Computing Machinery.

\bibitem[{Kardas et~al.(2020)Kardas, Czapla, Stenetorp, Ruder, Riedel, Taylor, and Stojnic}]{DBLP:journals/corr/abs-2004-14356}
Marcin Kardas, Piotr Czapla, Pontus Stenetorp, Sebastian Ruder, Sebastian Riedel, Ross Taylor, and Robert Stojnic. 2020.
\newblock \href {http://arxiv.org/abs/2004.14356} {Axcell: Automatic extraction of results from machine learning papers}.
\newblock \emph{CoRR}, abs/2004.14356.

\bibitem[{Kulal et~al.(2019)Kulal, Pasupat, Chandra, Lee, Padon, Aiken, and Liang}]{DBLP:journals/corr/abs-1906-04908}
Sumith Kulal, Panupong Pasupat, Kartik Chandra, Mina Lee, Oded Padon, Alex Aiken, and Percy Liang. 2019.
\newblock \href {http://arxiv.org/abs/1906.04908} {Spoc: Search-based pseudocode to code}.
\newblock \emph{CoRR}, abs/1906.04908.

\bibitem[{Lachaux et~al.(2020)Lachaux, Roziere, Chanussot, and Lample}]{https://doi.org/10.48550/arxiv.2006.03511}
Marie-Anne Lachaux, Baptiste Roziere, Lowik Chanussot, and Guillaume Lample. 2020.
\newblock \href {https://doi.org/10.48550/ARXIV.2006.03511} {Unsupervised translation of programming languages}.

\bibitem[{{LaTeXML Project}(2022)}]{latexml}
{LaTeXML Project}. 2022.
\newblock {LaTeXML}: A latex to xml/html/mathml converter.
\newblock \url{https://math.nist.gov/~BMiller/LaTeXML/}.

\bibitem[{Lu et~al.(2021)Lu, Guo, Ren, Huang, Svyatkovskiy, Blanco, Clement, Drain, Jiang, Tang, Li, Zhou, Shou, Zhou, Tufano, Gong, Zhou, Duan, Sundaresan, Deng, Fu, and Liu}]{https://doi.org/10.48550/arxiv.2102.04664}
Shuai Lu, Daya Guo, Shuo Ren, Junjie Huang, Alexey Svyatkovskiy, Ambrosio Blanco, Colin Clement, Dawn Drain, Daxin Jiang, Duyu Tang, Ge~Li, Lidong Zhou, Linjun Shou, Long Zhou, Michele Tufano, Ming Gong, Ming Zhou, Nan Duan, Neel Sundaresan, Shao~Kun Deng, Shengyu Fu, and Shujie Liu. 2021.
\newblock \href {https://doi.org/10.48550/ARXIV.2102.04664} {Codexglue: A machine learning benchmark dataset for code understanding and generation}.

\bibitem[{Mali et~al.(2020)Mali, Kukkadapu, Mahdavi, and Zanibbi}]{DBLP:journals/corr/abs-2003-08005}
Parag Mali, Puneeth Kukkadapu, Mahshad Mahdavi, and Richard Zanibbi. 2020.
\newblock \href {http://arxiv.org/abs/2003.08005} {Scanssd: Scanning single shot detector for mathematical formulas in {PDF} document images}.
\newblock \emph{CoRR}, abs/2003.08005.

\bibitem[{Mishra et~al.(2023)Mishra, Kumar, Bhat, Murthy, Contractor, and Tamilselvam}]{mishra-etal-2023-prompting}
Mayank Mishra, Prince Kumar, Riyaz Bhat, Rudra Murthy, Danish Contractor, and Srikanth Tamilselvam. 2023.
\newblock \href {https://aclanthology.org/2023.emnlp-main.939} {Prompting with pseudo-code instructions}.
\newblock In \emph{Proceedings of the 2023 Conference on Empirical Methods in Natural Language Processing}, pages 15178--15197, Singapore. Association for Computational Linguistics.

\bibitem[{Nassar et~al.(2022)Nassar, Livathinos, Lysak, and Staar}]{nassar2022tableformer}
Ahmed Nassar, Nikolaos Livathinos, Maksym Lysak, and Peter Staar. 2022.
\newblock \href {http://arxiv.org/abs/2203.01017} {Tableformer: Table structure understanding with transformers}.

\bibitem[{Oda et~al.(2015)Oda, Fudaba, Neubig, Hata, Sakti, Toda, and Nakamura}]{Oda2015LearningTG}
Yusuke Oda, Hiroyuki Fudaba, Graham Neubig, Hideaki Hata, Sakriani Sakti, Tomoki Toda, and Satoshi Nakamura. 2015.
\newblock \href {https://api.semanticscholar.org/CorpusID:15979705} {Learning to generate pseudo-code from source code using statistical machine translation (t)}.
\newblock \emph{2015 30th IEEE/ACM International Conference on Automated Software Engineering (ASE)}, pages 574--584.

\bibitem[{Odisho et~al.(2016)Odisho, Aziz, and Giacaman}]{odisho2016teaching}
Ogen Odisho, Mark Aziz, and Nasser Giacaman. 2016.
\newblock Teaching and learning data structure concepts via visual kinesthetic pseudocode with the aid of a constructively aligned app.
\newblock \emph{Computer Applications in Engineering Education}, 24(6):926--933.

\bibitem[{Peltsverger and Debnath(2019)}]{10.1145/3304221.3325581}
Svetlana Peltsverger and Sourav Debnath. 2019.
\newblock \href {https://doi.org/10.1145/3304221.3325581} {Instructional pseudocode guide to teach problem-solving}.
\newblock In \emph{Proceedings of the 2019 ACM Conference on Innovation and Technology in Computer Science Education}, ITiCSE '19, page 319, New York, NY, USA. Association for Computing Machinery.

\bibitem[{Roziere et~al.(2021)Roziere, Lachaux, Szafraniec, and Lample}]{https://doi.org/10.48550/arxiv.2102.07492}
Baptiste Roziere, Marie-Anne Lachaux, Marc Szafraniec, and Guillaume Lample. 2021.
\newblock \href {https://doi.org/10.48550/ARXIV.2102.07492} {Dobf: A deobfuscation pre-training objective for programming languages}.

\bibitem[{Rozi{\`e}re et~al.(2021)Rozi{\`e}re, Zhang, Charton, Harman, Synnaeve, and Lample}]{Rozire2021LeveragingAU}
Baptiste Rozi{\`e}re, J~Zhang, François Charton, Mark Harman, Gabriel Synnaeve, and Guillaume Lample. 2021.
\newblock Leveraging automated unit tests for unsupervised code translation.
\newblock \emph{ArXiv}, abs/2110.06773.

\bibitem[{Shi et~al.(2020)Shi, Bieber, and Sutton}]{pmlr-v119-shi20a}
Kensen Shi, David Bieber, and Charles Sutton. 2020.
\newblock \href {https://proceedings.mlr.press/v119/shi20a.html} {Incremental sampling without replacement for sequence models}.
\newblock In \emph{Proceedings of the 37th International Conference on Machine Learning}, volume 119 of \emph{Proceedings of Machine Learning Research}, pages 8785--8795. PMLR.

\bibitem[{Sontakke et~al.(2023)Sontakke, Kalra, Patwardhan, Vig, Medicherla, Naik, and Pradhan}]{sontakke2023knowledge}
Ankita Sontakke, Kanika Kalra, Manasi Patwardhan, Lovekesh Vig, Raveendra~Kumar Medicherla, Ravindra Naik, and Shrishti Pradhan. 2023.
\newblock \href {http://arxiv.org/abs/2303.09062} {Knowledge transfer for pseudo-code generation from low resource programming language}.

\bibitem[{{The Apache Software Foundation}(2012)}]{pdfbox2012}
{The Apache Software Foundation}. 2012.
\newblock \href {http://pdfbox.apache.org/} {Apache pdfbox projekt}.

\bibitem[{Xie et~al.(2021)Xie, Ma, and Liang}]{pmlr-v139-xie21f}
Sang~Michael Xie, Tengyu Ma, and Percy Liang. 2021.
\newblock \href {https://proceedings.mlr.press/v139/xie21f.html} {Composed fine-tuning: Freezing pre-trained denoising autoencoders for improved generalization}.
\newblock In \emph{Proceedings of the 38th International Conference on Machine Learning}, volume 139 of \emph{Proceedings of Machine Learning Research}, pages 11424--11435. PMLR.

\bibitem[{Xu et~al.(2022)Xu, Alon, Neubig, and Hellendoorn}]{xu2022systematic}
Frank~F. Xu, Uri Alon, Graham Neubig, and Vincent~J. Hellendoorn. 2022.
\newblock \href {http://arxiv.org/abs/2202.13169} {A systematic evaluation of large language models of code}.

\bibitem[{Yang et~al.(2021)Yang, Zhou, Chen, and Yu}]{Yang2021FinegrainedPG}
Guang Yang, Yanlin Zhou, Xiang Chen, and Chi Yu. 2021.
\newblock \href {https://api.semanticscholar.org/CorpusID:236139947} {Fine-grained pseudo-code generation method via code feature extraction and transformer}.
\newblock \emph{2021 28th Asia-Pacific Software Engineering Conference (APSEC)}, pages 213--222.

\bibitem[{Yasunaga and Liang(2020)}]{10.5555/3524938.3525939}
Michihiro Yasunaga and Percy Liang. 2020.
\newblock Graph-based, self-supervised program repair from diagnostic feedback.
\newblock In \emph{Proceedings of the 37th International Conference on Machine Learning}, ICML'20. JMLR.org.

\bibitem[{Zavershynskyi et~al.(2018)Zavershynskyi, Skidanov, and Polosukhin}]{zavershynskyi2018naps}
Maksym Zavershynskyi, Alex Skidanov, and Illia Polosukhin. 2018.
\newblock \href {http://arxiv.org/abs/1807.03168} {Naps: Natural program synthesis dataset}.

\bibitem[{Zhang et~al.(2022)Zhang, Xu, Xiao, and Xue}]{unipseudo}
Weiwei Zhang, Zhengzi Xu, Yang Xiao, and Yinxing Xue. 2022.
\newblock \href {https://doi.org/10.1186/s42400-022-00121-0} {Unleashing the power of pseudo-code for binary code similarity analysis}.
\newblock \emph{Cybersecurity}, 5.

\bibitem[{Zhong et~al.(2020)Zhong, Stern, and Klein}]{DBLP:journals/corr/abs-2005-05927}
Ruiqi Zhong, Mitchell Stern, and Dan Klein. 2020.
\newblock \href {http://arxiv.org/abs/2005.05927} {Semantic scaffolds for pseudocode-to-code generation}.
\newblock \emph{CoRR}, abs/2005.05927.

\end{thebibliography}

\end{document}